\documentclass[twocolumn,pra,twocolumn,showpacs]{revtex4}
\usepackage[T1]{fontenc}
\usepackage[latin9]{inputenc}
\usepackage{amsmath}
\usepackage{amssymb}
\usepackage{graphicx}
\usepackage{esint}

\makeatletter

\@ifundefined{textcolor}{}
{%
 \definecolor{BLACK}{gray}{0}
 \definecolor{WHITE}{gray}{1}
 \definecolor{RED}{rgb}{1,0,0}
 \definecolor{GREEN}{rgb}{0,1,0}
 \definecolor{BLUE}{rgb}{0,0,1}
 \definecolor{CYAN}{cmyk}{1,0,0,0}
 \definecolor{MAGENTA}{cmyk}{0,1,0,0}
 \definecolor{YELLOW}{cmyk}{0,0,1,0}
 }

\@ifundefined{textcolor}{}{%
 \definecolor{BLACK}{gray}{0}
 \definecolor{WHITE}{gray}{1}
 \definecolor{RED}{rgb}{1,0,0}
 \definecolor{GREEN}{rgb}{0,1,0}
 \definecolor{BLUE}{rgb}{0,0,1}
 \definecolor{CYAN}{cmyk}{1,0,0,0}
 \definecolor{MAGENTA}{cmyk}{0,1,0,0}
 \definecolor{YELLOW}{cmyk}{0,0,1,0}
 }

\makeatother

\begin{document}

\title{Waveguide QED: controllable channel from quantum interference}

\author{Qiong Li$^{1}$, Lan Zhou$^{2}$, and C. P. Sun$^{3}$}

\email{cpsun@csrc.ac.cn}

\homepage{http://www.csrc.ac.cn/ suncp/}

\affiliation{$^{1}$State Key Laboratory of Theoretical Physics,
Institute of Theoretical Physics, University of Chinese Academy of
Science, Beijing
100190, China\\
 $^{2}$Department of Physics, Hunan Normal University, Changsha 410081,
China\\
 $^{3}$Beijing Computational Science Research Center, Beijing 100084,
China}

\begin{abstract}
We study a waveguide QED system with a rectangular waveguide and a
two-level system (TLS) inside, where the transverse modes TM$_{mn}$
define the quantum channels of guided photons. It is discovered that
the loss of photons in the TM$_{11}$ channel into the others can be
overcome by replacing it with a certain coherent superposition of
TM$_{mn}$ channels, which is named as the controllable channel (CC)
as the photons in CC can be perfectly reflected or transmitted
 by the TLS, and never lost into the other channels. The dark state
emerges when the photon is incident from one of the scattering-free
channels (SFCs) orthogonal to CC. The underlying physics mechanism
is the multi-channel interference associated with Fano resonance.
\end{abstract}


\maketitle

In a fully-quantum network based on single photon carriers to
process quantum information, the essential task is to coherently
control photon propagation by a local quantum node
\cite{key1-Harris1998,key2-Ham2000,key3-Birnbaum2005,key4-Bajcsy2009,key5-Shen,ShenPRA,Hu2007,key6-Zhou2008,key7-Gong2008,ShiPRB,LiaoPRA,AMO2010,key8-Chang2007,key9-Hwang2009,key10-Tsoi2009,key11-Shi2011,key12-Zhang2012,key13-Wang2012,key14-Aoki2009,key15-Zhou-arxiv}.
To this end, a hybrid system consisting of a one-dimensional (1D)
waveguide coupled to a two-level system (TLS) is extensively studied
for physical implementation of the quantum node acting as a quantum
switch
\cite{key4-Bajcsy2009,key5-Shen,ShenPRA,Hu2007,key6-Zhou2008,key7-Gong2008,ShiPRB,LiaoPRA,AMO2010}
or a single photon transistor \cite{key8-Chang2007,key9-Hwang2009}.
With the single mode approximation for a waveguide with
infinitesimal cross section, the total reflection of single photons
by the TLS was found to be responsible for the dominant functions of
quantum devices.

However, a realistic waveguide with a finite cross section
necessarily possesses transverse modes. Thus, photons guided in the
realistic waveguide may be in different quantum channels defined by
transverse modes. Each transverse mode has a cut-off frequency for
the corresponding guiding mode. To demonstrate multi-channel effects
on single-photon scattering, Ref.\cite{Huang-arxiv} made an
approximation using two modes and a quadratic dispersion relation,
and showed that the guided photon cannot be totally reflected due to
the loss from one mode to the other. In other words, the quantum
device oriented functions we desire can not be well achieved . In
order to overcome such multi-channel loss, we will revisit the
waveguide QED by considering a realistic hybrid system without any
over-approximation.

In this letter, we study single-photon scattering by a TLS locally
embedded in a waveguide of finite rectangular cross section. In our
approach, both the real dispersion relation and multi- channel
effects are exactly taken into account. As for the multi-channel
induced loss, we find that there exists a unique controllable
channel (CC) defined by a particular superposition of the transverse
magnetic modes TM$_{mn}$, in which the guided photons can be well
controlled by the TLS since the guided photons in the complementary
channels orthogonal to CC are completely decoupled from the TLS.
Such a scattering free channel (SFC) behaves as the dark state to
support the electromagnetically induced transparency (EIT). The
controllable channel and all SFCs make up the whole
single-excitation Hilbert space of the waveguide QED system. In the
controllable channel the quantum interference between the incident
wave and the scattered one leads to a Fano resonance\cite{Fano}, so
that the photon could be perfectly reflected or enabled to be
completely transmitted by the TLS. Therefore, using the CC to guide
photons we can well exploit the quantum device oriented functions of
the TLS in a realistic waveguide.

\begin{figure}
\includegraphics[bb=270 250 490 415, width=6cm]{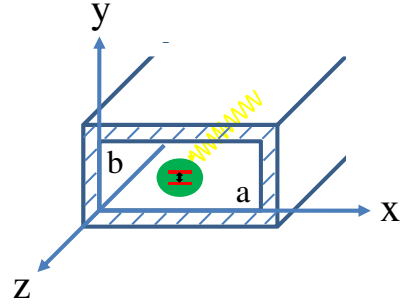}
\caption{\label{fig:setup} (Color online) A rectangular waveguide
with a TLS located in the center of the cross section. The TLS is
dipole-coupled to the TM guiding modes of the waveguide. $a=2b$.}
\end{figure}

\begin{figure}
\includegraphics[bb=100 220 490 665, width=6cm]{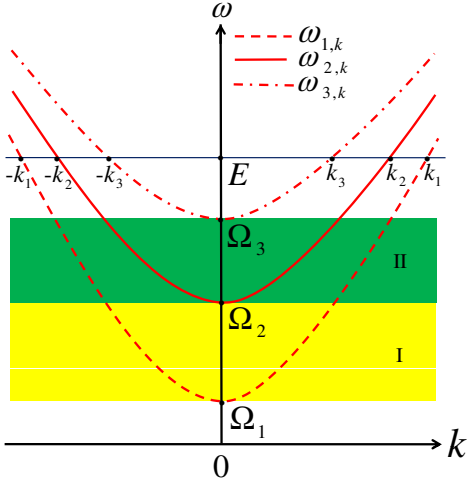}
\caption{\label{fig:dispersion} The dispersion relation of the TM
guiding modes of the waveguide. $\Omega_{1}=2.23607$,
$\Omega_{2}=3.60555$, $\Omega_{3}=6.08276$. The energy is in the
unit of $\pi/a$. The photon with energy $E>\Omega_{3}$ can propagate
in different modes with the wavenumber $k=k_{1}$, $k_{2}$, $k_{3}$
etc. }
\end{figure}

\textit{Photon scattering by a TLS within a rectangular
waveguide}.We consider a waveguide of rectangular cross section with
area $A=ab$, as shown in Fig.\ref{fig:setup}. The guiding modes in
such a realistic waveguide are labeled by $(m,n,k)$, with $(m,n)$
the transverse magnetic mode TM$_{mn}$ (standing wave-numbers in the
cross section are $k_{x}=m\pi/a$ , $k_{y}=n\pi/b$) and $k$ the
propagating wavenumber along the $z$-direction. Each transverse mode
$(m,n)$ has the cut-off frequency
$\Omega_{mn}=\sqrt{\left(m\pi/a\right)^{2}+\left(n\pi/b\right)^{2}}$
(the unit $\hbar=1=c$ is used). According to the ascending order of
the cut-off frequencies, we replace $(m,n)$ with its sequence number
$j$ , that is $j=1,2,3.....$ denote the transverse modes TM$_{11}$,
TM$_{31}$,TM$_{13}$.... respectively. The dispersion relation of the
guiding modes is given by $\omega_{j,k}=\sqrt{\Omega_{j}^{2}+k^{2}}$
, as plotted in Fig.\ref{fig:dispersion}. The TLS of transition
frequency $\omega_{a}$ is located at
$\mathbf{r}_{a}=\left(a/2,b/2,0\right)$, whose ground (excited)
state is denoted by $|g\rangle$ ($|e\rangle$). The dipole oriented
along the z-direction couples the TM electric field. With the atomic
rising (lowering) operator
$\sigma_{+}(\sigma_{-})\equiv|e\rangle\langle g|$ ($|g\rangle\langle
e|$), the total Hamiltonian $H=H_{0}+V$ of the hybrid system is
given by the free part
\begin{equation}
H_{0}=\sum_{j}\int_{-\infty}^{+\infty}dk\omega_{j,k}a_{j,k}^{\dagger}a_{j,k}+\omega_{a}\sigma_{+}\sigma_{-},\label{EqII-1a}
\end{equation}
 and the dipole interaction
\begin{equation}
V=\sum_{j}\int_{-\infty}^{+\infty}dk\left(g_{j,k}^{*}a_{j,k}\sigma_{+}+h.c.\right).\label{EqII-1b}
\end{equation}
Here, the mode-dependent coupling strength reads
\begin{equation}
g_{j,k}=-\frac{g\Omega_{j}}{\sqrt{\omega_{j,k}}}\sin\frac{m\pi}{2}\sin\frac{n\pi}{2}\label{EqII-2}
\end{equation}
with $j\equiv(m,n)$, $g=d/\sqrt{\pi A}$. The matrix element $d$ of
the dipole transition is set to be real. Note that the coupling
strength vanishes for even integers $m$ or $n$.

\begin{figure}
\includegraphics[bb=205 182 680 420, width=7cm]{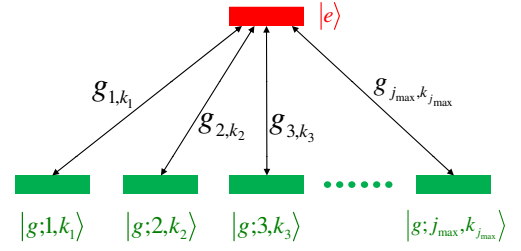}
\caption{\label{fig:dark} (Color online) The schematic illustration of the
TLS as a multi-component dark state. }
\end{figure}

We now consider single-photon scattering in the waveguide QED
system. For a single photon initially in the state $\left\vert
\phi_{\text{in}}\right\rangle
=a_{j,k}^{\dagger}\left|\varnothing\right\rangle $ with energy
$E=\omega_{j,k}$, the scattering state assumes the form
\begin{align}
\left\vert \phi^{(+)}\right\rangle  & =\sum_{j'}\int dk'U_{g}(j',k';j,k)a_{j',k'}^{\dagger}\left\vert \varnothing\right\rangle \nonumber \\
 & +U_{e}(j,k)\sigma_{+}\left\vert \varnothing\right\rangle ,\label{eq:scattering}
\end{align}
which ensures the conservation of the excitation number in the
single excitation subspace. Here, $\left|\varnothing\right\rangle $
represents the TLS in the ground state and the waveguide field in
vacuum. The amplitudes $U_{g}$ and $U_{e}$ are obtained from the
Lippmann Schwinger equation $\left\vert \phi^{(+)}\right\rangle
=\left\vert \phi_{\text{in}}\right\rangle
+(E-H_{0}+i0^{+})^{-1}V\left\vert \phi^{(+)}\right\rangle .$ The
elements of the scattering matrix $\hat{S}$ can be obtained from the
scattering state as
\begin{align}
\left\langle j',k'\right|\hat{S}\left|j,k\right\rangle  & =\delta_{j,j'}\delta(k-k')-2\pi i\delta\left(\omega_{j',k'}-\omega_{j,k}\right)\nonumber \\
 & \times\frac{g_{j^{\prime},k^{\prime}}^{*}g_{j,k}}{E-\omega_{a}-\Delta(E)+i\Gamma(E)},\label{eq:S-matrix}
\end{align}
where $\left|j,k\right\rangle \equiv
a_{j,k}^{\dagger}\left|\varnothing\right\rangle $,
$\Delta(E)=\sum_{j}\int_{-\infty}^{+\infty}dk\mathcal{P}\left(1/(E-\omega_{j,k})\right)\left\vert
g_{j,k}\right\vert ^{2}$and
$\Gamma(E)=\pi\sum_{j}\int_{-\infty}^{+\infty}dk\delta(E-\omega_{j,k})g_{j,k}^{*}g_{j,k}$
are the real and imaginary parts of the self-energy
$\Sigma(E)=\Delta(E)-i\Gamma(E)$ respectively. For detailed
calculations of the scattering matrix element, see the supplemental
material.

\textit{Single channel scattering and its loss}. It is known that a
TLS acts as a quantum switch for single photons confined in a
single-mode waveguide
\cite{key5-Shen,key6-Zhou2008,key7-Gong2008,key8-Chang2007}. To keep
a photon propagating in a single quantum channel, this realistic
waveguide is required that: 1) The cross section must be so small
that $\Omega_{2}-\Omega_{1}$ is large enough; 2) The energy of the
input photon $\omega_{1,k}$ is below the cutoff frequency
$\Omega_{2}$, i.e. $k<\sqrt{\Omega_{2}^{2}-\Omega_{1}^{2}}$. Under
these conditions, Eq.\eqref{eq:S-matrix} gives the reflection
amplitude in the TM$_{11}$ mode as
\begin{eqnarray}
r & = &
\frac{-i\Gamma(\omega_{1,k})}{\omega_{1,k}-\omega_{a}-\Delta(\omega_{1,k})+i\Gamma(\omega_{1,k})}.
\end{eqnarray}
Obviously, the total reflection occurs in the resonance condition
$\omega_{1,k}-\omega_{a}-\Delta(\omega_{1,k})=0$, which becomes
$\omega_{1,k}\simeq\omega_{a}+\Delta(\omega_{a})\equiv\omega_{A}$ in
the weak coupling limit. Note that in previous studies
\cite{key5-Shen,key6-Zhou2008,key7-Gong2008,key8-Chang2007,Huang-arxiv},
the Lamb shift $\Delta(\omega_{a})$, which arises from the
renormalization of the TLS's energy level, has been ignored due to
the use of the quadratic or linear dispersion approximation.

\begin{figure}
\includegraphics[bb=18 8 362 265, width=6cm]{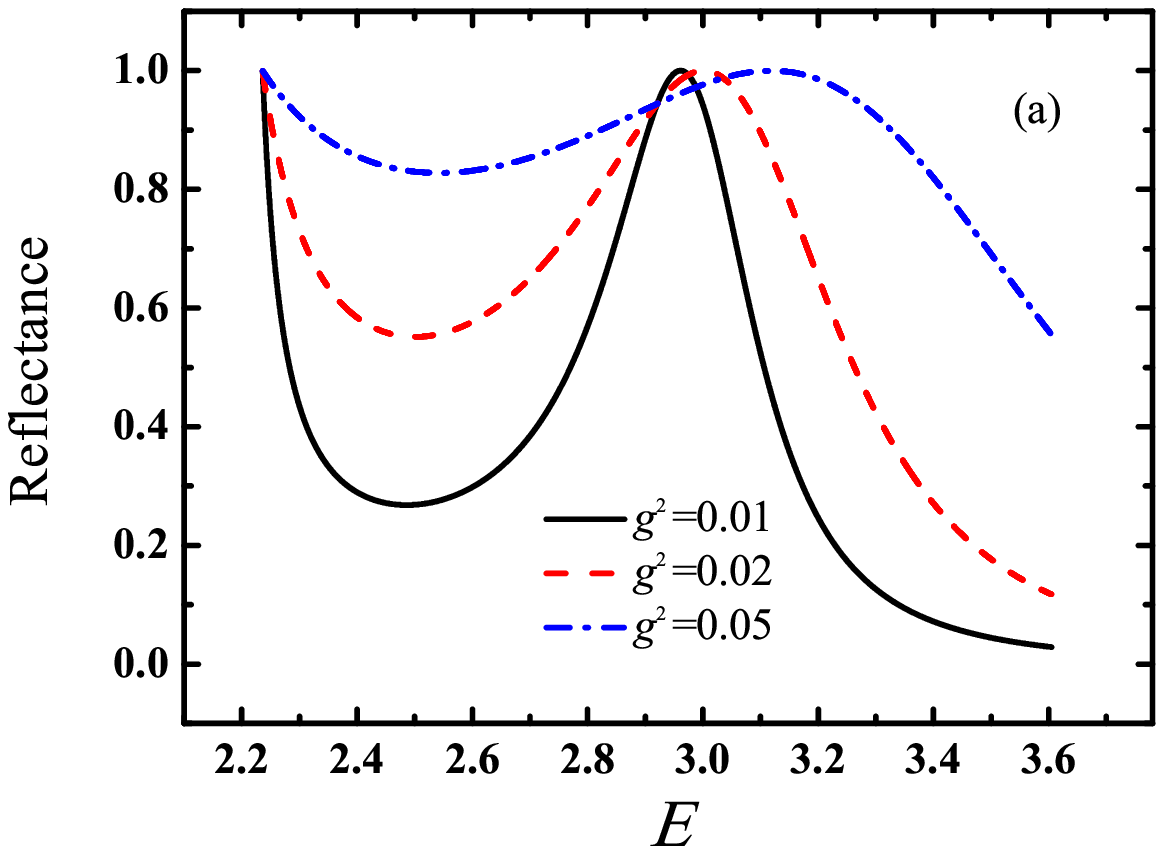}
\\
\includegraphics[bb=18 8 362 265, width=6cm]{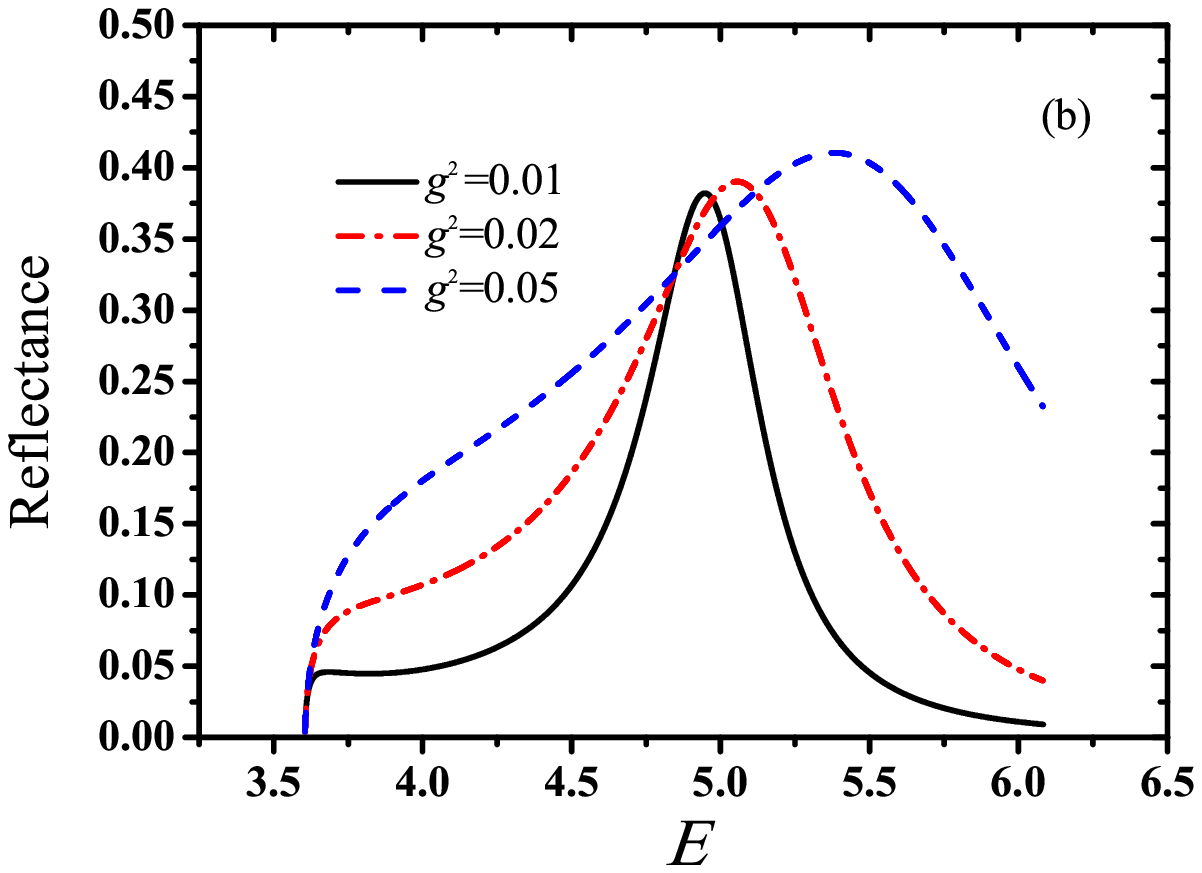}
\caption{\label{fig:single-loss}(Color online) The reflectance
spectrum for a single photon incident in the mode TM$_{11}$. (a)
$E\in(\Omega_{1},\Omega_{2})$,
$\omega_{a}=(\Omega_{1}+\Omega_{2})/2$. (b)
$E\in(\Omega_{2},\Omega_{3})$,
$\omega_{a}=(\Omega_{2}+\Omega_{3})/2$. The energy $E$ is in the
unit of $\pi/a$.}
\end{figure}

\begin{figure}
\includegraphics[bb=18 8 362 265, width=6cm]{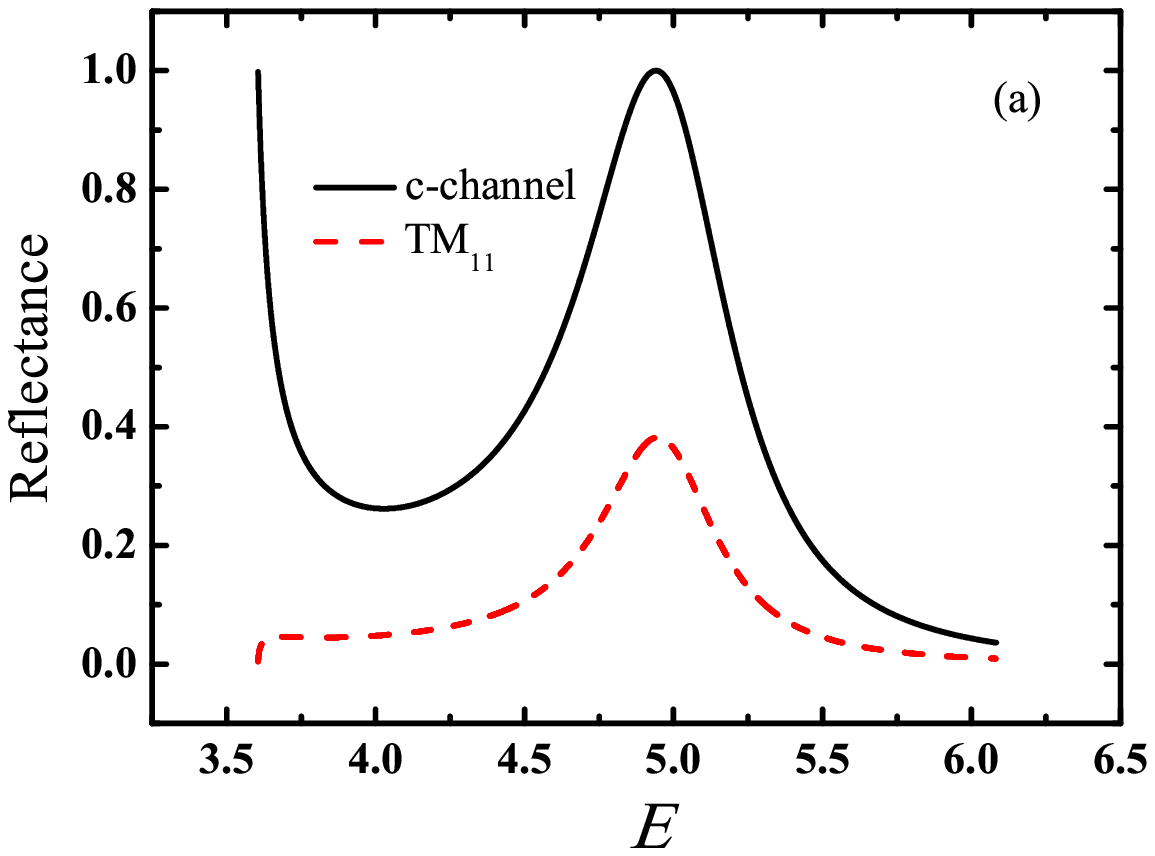}
\\
\includegraphics[bb=18 8 362 265, width=6cm]{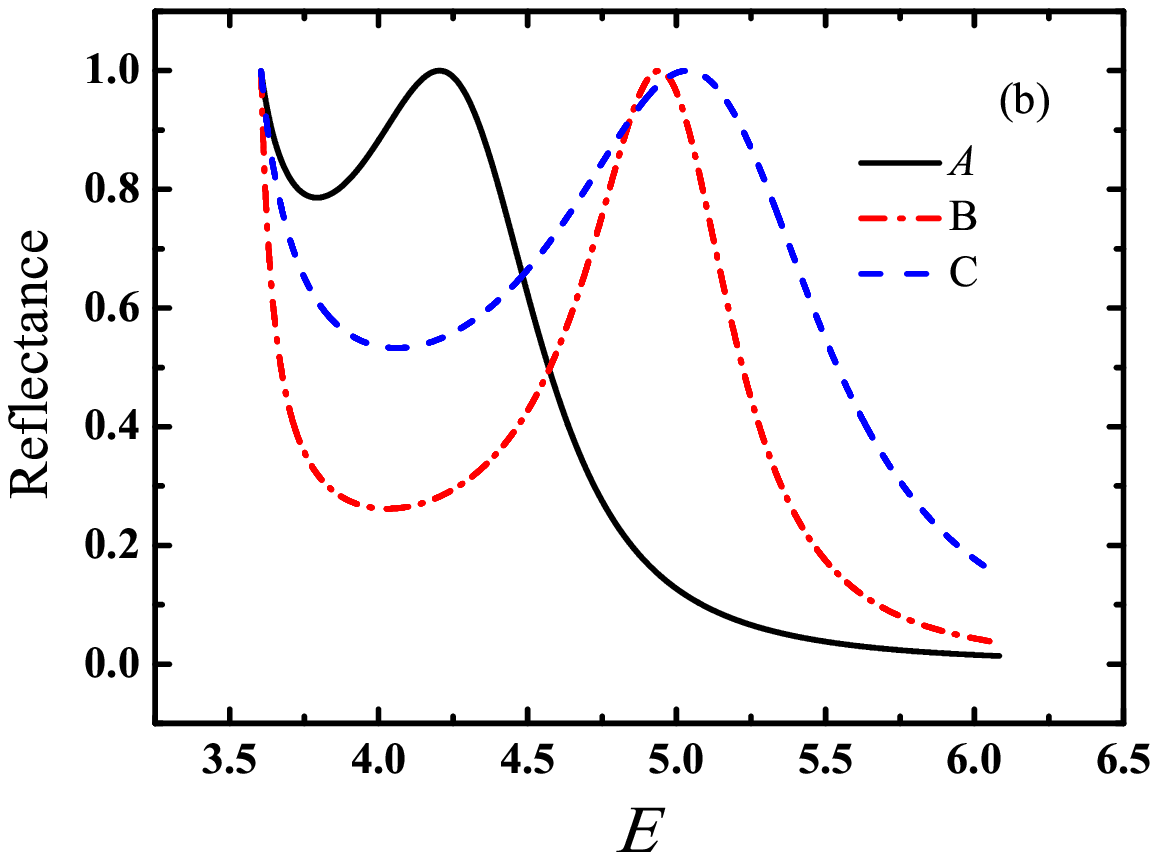}
\caption{\label{fig:control} (Color online) (a) The reflectance
spectrum with $E\in(\Omega_{2},\Omega_{3})$. The single photon is
input in the mode TM$_{11}$ (red dashed line) or input in the
controllable channel (black solid line). $g^{2}=0.01$.
$\omega_{a}=(\Omega_{2}+\Omega_{3})/2$. (b) The reflectance spectrum
with $E\in(\Omega_{2},\Omega_{3})$ for a single photon input in the
controllable channel. A: $\omega_{a}=0.8\Omega_{2}+0.2\Omega_{3}$,
$g^{2}=0.01$; B: $\omega_{a}=0.5(\Omega_{2}+\Omega_{3})$,
$g^{2}=0.01$; C: $\omega_{a}=0.5(\Omega_{2}+\Omega_{3})$,
$g^{2}=0.02$. The energy $E$ is in the unit of $\pi/a$.}
\end{figure}

\begin{figure}
\includegraphics[bb=18 8 362 265, width=6cm]{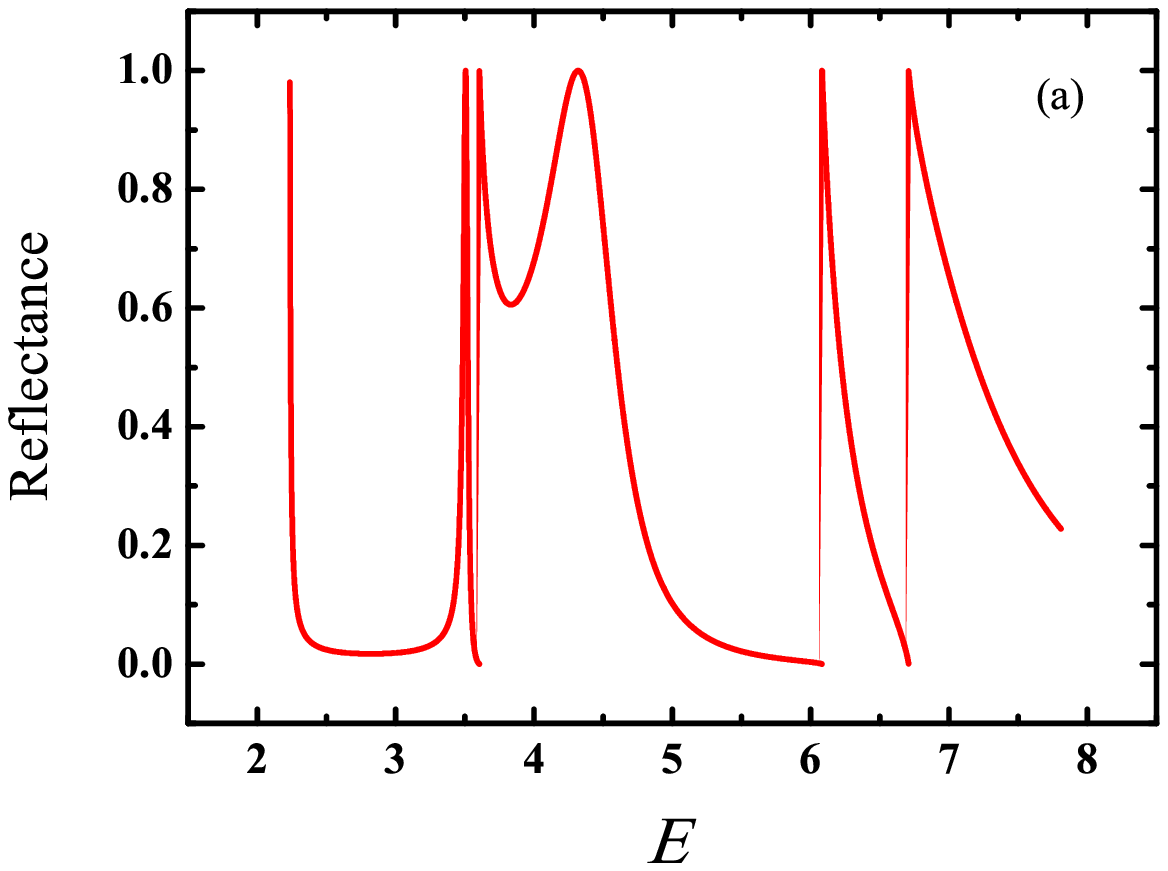}
\\
\includegraphics[bb=18 8 362 265, width=6cm]{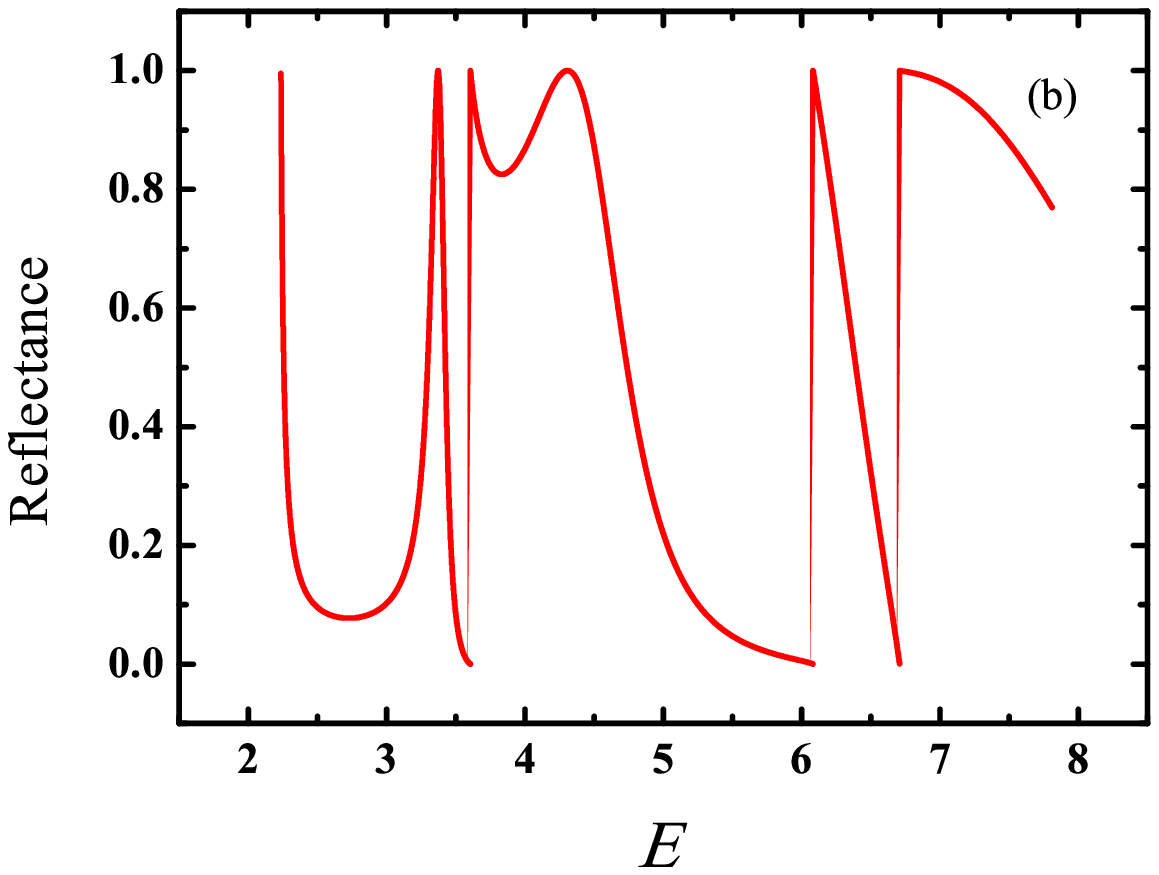}
\caption{\label{fig:multi-peak} (Color online) The reflectance
spectrum with $E\in(\Omega_{1},\Omega_{5})$ for a single photon
input in the controllable channel. $\Omega_{1}=2.236$,
$\Omega_{2}=3.605$, $\Omega_{3}=6.082$, $\Omega_{4}=6.708$,
$\Omega_{5}=7.810$. (a) $g^{2}=0.01$.
$\omega_{a}=(\Omega_{1}+\Omega_{5})/2$. (b) $g^{2}=0.02$.
$\omega_{a}=(\Omega_{2}+\Omega_{5})/2$. The energy $E$ is in the
unit of $\pi/a$.}
\end{figure}

In Fig.\ref{fig:single-loss}, we plot the reflectance $R=|r|^{2}$ as
a function of the incident energy $E=\omega_{1,k}$. A single photon
confined in the TM$_{11}$ mode (see Fig.\ref{fig:single-loss}(a)) is
indeed perfectly reflected by the TLS provided that $\omega_{a}$ is
in the central domain of the range
$\left[\Omega_{1},\Omega_{2}\right]$. However, the position of the
total reflection experiences a blue shift to $\omega_{a}$ due to the
renormalization, which becomes larger as the coupling strength
increases. We note that total reflection also occurs when
$E\rightarrow\Omega_{1}+0^{+}$, which is referred to as the cut-off
frequency resonance \cite{Huang-arxiv}. When the incident energy $E$
is above $\Omega_{2}$, higher-order modes and the induced
multi-channel interference effects must be taken into account. The
reflectance with $E\in\left[\Omega_{2},\Omega_{3}\right]$ is plotted
in Fig. \ref{fig:single-loss}(b). Although $\omega_{A}$ still
determines the reflection peak, the maximum becomes smaller than
unity, showing that single photons in the TM$_{11}$ mode experience
a finite loss due to the existence of the higher-order modes in a
realistic waveguide. Actually, this loss is caused by the TLS
mediating the resonant tunneling process between the TM$_{11}$ mode
and higher-order modes.

\textit{Controllable channel versus scattering-free channels}. Now,
we study the quantum interference among different TM$_{mn}$ modes.
We assume that a single photon with energy $E$ incident from the
negative z-direction is initially in this superposition state
\begin{equation}
\left\vert \phi_{\text{in}}\right\rangle
=\sum_{j=1}^{j_{\text{max}}(E)}\varphi_{j}\left\vert
j,k_{j}\right\rangle ,
\end{equation}
where $j_{\text{max}}(E)$ is the highest mode a photon with energy
$E$ can reach, fixed by the condition
$\Omega_{j_{\text{max}}(E)}<E<\Omega_{j_{\text{max}}(E)+1}$. The
complex coefficients $\varphi_{j}$ (with $j=1,2,3..$) represent the
amplitudes in the $j$th mode. Here,
$k_{j}=\sqrt{E^{2}-\Omega_{j}^{2}}(j=1,2,3\cdots)$ takes the
discrete values illustrated in Fig.\ref{fig:dispersion}, i.e., the
crossing points between the horizontal line $\omega_{j,k}=E$ and the
dispersion curves. With the scattering matrix elements in
Eq.\eqref{eq:S-matrix}, the multi-channel outgoing state $\left\vert
\phi_{\text{out}}\right\rangle \equiv\hat{S}\left\vert
\phi_{\text{in}}\right\rangle $ reads
\begin{eqnarray}
\left|\phi_{\text{out}}\right\rangle  & = & \left\vert \phi_{\text{in}}\right\rangle -2\pi i\sum_{j'}\sqrt{\rho_{j'}(E)}\varphi_{j'}g_{j',k_{j'}}^{*}\nonumber \\
 &  & \times\frac{\sum_{j}\sqrt{\rho_{j}(E)}g_{j,k_{j}}}{E-\omega_{a}-\Delta(E)+i\Gamma(E)}\left(a_{j,k_{j}}^{\dagger}+a_{j,-k_{j}}^{\dagger}\right)\left|\varnothing\right\rangle ,\nonumber \\
\label{eq:phiout}
\end{eqnarray}
where $\rho_{j}(E)\equiv E/\sqrt{E^{2}-\Omega_{j}^{2}}$,
$a_{j,k_{j}}^{\dagger}$ and $a_{j,-k_{j}}^{\dagger}$ are the right-
and left-going creation operators respectively. For a photon with
energy $E$, the Hilbert space can be decomposed as the vector
$\varphi^{(c)}$, which is proportional to the vector
$\mathbf{g}\sqrt{\rho(E)}\equiv\left(g_{1,k_{1}}\sqrt{\rho_{1}(E)},g_{2,k_{2}}\sqrt{\rho_{2}(E)},\cdots,g_{j_{\text{max}},k_{j_{\text{max}}}}\sqrt{\rho_{j_{\text{max}}}(E)}\right)$,
and the complementary subspace, which is spanned by the vectors
$\varphi^{(F)}$ orthogonal to $\mathbf{g}\sqrt{\rho(E)}$, i.e.
$\sum_{j=1}^{j_{\text{max}}(E)}\sqrt{\rho_{j}(E)}g_{j,k_{j}}^{*}\varphi_{j}^{(F)}=0.$
Obviously, single photons incident in the states $\varphi^{(F)}$
from one side of the rectangular waveguide will be completely
transmitted to the other side. Thus, the vectors $\varphi^{(F)}$
span a so-called scattering-free subspace to define the SFCs where
the confined photons are decoupled from the TLS. This phenomenon is
similar to the dark state of a three-level atom to support the
electromagnetic induce transparency (EIT). Actually, the
scattering-free state
$|\varphi^{(F)}\rangle=\sum_{j=1}^{j_{\text{max}}(E)}\varphi_{j}^{(F)}|j,k_{j}\rangle$(with
the photon energy $E$) can be understood as a multi-component dark
state as illustrated in Fig.\ref{fig:dark}. There exist the
multi-channel transitions from $|g;j,k_{j}\rangle(j=1,2,\cdots
j_{max})$ to $|e\rangle$. The quantum interference among these
channels results in the transparency of the TLS with respect to the
incident photon. Thus, no reflected photon is observed.

The remaining vector orthogonal to the scattering-free subspace,
defined by $\varphi_{j}^{(c)}\propto g_{j,k_{j}}\sqrt{\rho_{j}(E)}$,
is regarded as the controllable channel. In this channel, the TLS
scatters the incident wave coming from the right into a
superposition of the right- and left-going waves, which we denote as
$\left|\phi_{\text{out}}\right\rangle
=\sum_{j=1}^{j_{\text{max}}(E)}\left(\varPhi_{j}^{(R)}a_{j,k_{j}}^{\dagger}+\varPhi_{j}^{(L)}a_{j,-k_{j}}^{\dagger}\right)\left\vert
\varnothing\right\rangle $. From Eq.\eqref{eq:phiout} the right- and
left-going wave function is obtained as
\begin{eqnarray}
\varPhi_{j}^{(R)} & = & \frac{E-\omega_{a}-\Delta(E)}{E-\omega_{a}-\Delta(E)+i\Gamma(E)}\varphi_{j}^{(c)},\label{eq:R}\\
\varPhi_{j}^{(L)} & = &
\frac{-i\Gamma(E)}{E-\omega_{a}-\Delta(E)+i\Gamma(E)}\varphi_{j}^{(c)},\label{eq:L}
\end{eqnarray}
which is proportional to the incident wave function
$\varphi_{j}^{(c)}$. The summation $\left\vert
\varPhi_{j}^{(R)}\right\vert ^{2}+\left\vert
\varPhi_{j}^{(L)}\right\vert ^{2}=\left\vert
\varphi_{j}^{(c)}\right\vert ^{2}$ guarantees that there is no loss
in the scattering process. Furthermore, in the weak coupling limit,
one can observe a perfect reflection once the incident energy $E$
matches the renormalized energy $\omega_{A}$ of the TLS with the
Lamb shift $\Delta(\omega_{a})$. As the cross section of the
waveguide increases, the dimension of the scattering-free subspace
increases for a given incident energy, but there still exists one
controllable channel. Therefore, total reflection can always be
observed as long as the incident photon is prepared in the
controllable channel, regardless of how large the cross section is.

When the incident energy approaches any cut-off frequency
$E\rightarrow\Omega_{j}+0^{+}$, $\Gamma(E)\rightarrow\infty$ leads
to $\varPhi_{j}^{(R)}=0$, and
$\varPhi_{j}^{(L)}=-\varphi_{j}^{(c)}$. Consequently, single photons
in the controllable channel can also be totally reflected when the
energy of the photon matches the cut-off frequency $\Omega_{j}$ of
any TM$_{mn}$ mode. We refer to the total reflection due to
$\Gamma(E)\rightarrow\infty$ as the cut-off frequency resonance
\cite{Huang-arxiv}.

In Fig.\ref{fig:control}, the reflectance spectrum is numerically
plotted for single photons incident in the controllable channel or
in the TM$_{11}$ mode. It can be found from Fig.\ref{fig:control}(a)
that when higher-order modes are taken into account, it is
impossible to find a constructive interference between the incoming
wave and the spontaneous emission from the TLS in a given mode.
However, single photons spontaneously emitted by the TLS are
confined to the controllable channel. Consequently, spontaneous
emission of the excited TLS can be exploited to control the coherent
transport properties of single photons in the controllable channel,
as shown in Fig.\ref{fig:control}(b). And the position of the total
reflection is shifted from the atomic transition frequency due to
the Lamb shift.

If the TLS-waveguide coupling is too strong, the resonance energy
$E_{R}$ can only be determined by solving the transcendental
equation $E_{R}-\omega_{a}-\Delta(E_{R})=0$, which may have more
than one solutions, rather than the single solution
$E_{R}\simeq\omega_{A}=\omega_{a}+\Delta(\omega_{a})$ in the weak
coupling limit. Here, we consider the contribution of several low
modes to the Lamb shift and a precise calculation shall be presented
elsewhere. By including more modes, multi-peaks are observed in the
reflectance spectrum as illustrated in Fig.\ref{fig:multi-peak}.

\textit{Conclusion.} We have carried out a systematic study about
the coherent scattering of single photons by a TLS in the realistic
rectangular waveguide. Usually, the dipole oriented along the
$z-$direction may scatter single photons guided in a TM$_{mn}$ mode
into the others, but a photon with energy $E<\Omega_{2}$ incident in
the TM$_{11}$mode can still be confined in TM$_{11}$ after
scattering. In this case, the TLS acts as an ideal quantum switch
when the renormalized frequency $\omega_{A}$ is in the range
$(\Omega_{1},\Omega_{2})$. However, with higher energy
$E>\Omega_{2}$, the single photons incident in TM$_{11}$ resonantly
tunnel to higher-order modes via the TLS.

For an artificial atom of transition frequency
$\omega_{A}\simeq10.2$GHz \cite{key-23} to work as a quantum switch,
the two lowest cut-off frequencies of the waveguide should satisfy
$\omega_{A}\simeq\frac{1}{2}\left(\Omega_{1}+\Omega_{2}\right)$,
leading to the size of the cross section $a/2=b\simeq2.1cm$.
Correspondingly, $\Omega_{2}\simeq79.1$GHz, so that the scattering
of microwave photons with energy $E\gtrsim79.1$GHz cannot be
confined in the single mode TM$_{11}$ and thus the multi-channel
effects (loss and interference) are involved. If we want to control
photons with higher energy, e.g. with energy about
$\omega_{A}\simeq1000$GHz, then we should use a waveguide of the
size $a/2=b\simeq2.1\mu m$ to work in the single mode region.
Conversely, if $b>2.7\mu m$, i.e. $\Omega_{2}<1000$GHz, then the
waveguide works in the multi-channel region, and in consequence we
must utilize the controllable channel scheme to overcome the channel
loss. The existence of the unique TLS-controllable channel and the
complementary scattering-free channels guarantees the success in
controlling single photons in a realistic waveguide with a finite
cross section.

We are grateful to C. Y. Cai, T. Tian, J. F. Huang, Y. Li, P. Zhang
and D. Z. Xu for helpful discussions. This work is supported by
National Natural Science Foundation of China under Grants
No.11121403, No.10935010, No.11222430, No.11074305, No. 11074261,
No. 11074071 and National 973 program under Grants No. 2012CB922104,
No. 2012CB922103. Hunan Provincial Natural Science Foundation of China (12JJ1002).

\end{document}